\documentclass[12pt]{article}
\usepackage{enumerate}
\usepackage{url} %

\usepackage[table]{xcolor}
\usepackage{soul}
\usepackage{comment}
\usepackage{units}
\usepackage{subcaption}
\usepackage{amsthm,amsfonts}
\usepackage{dsfont}
\usepackage{amssymb,amsmath,mathtools,amsfonts}
\usepackage{nccmath}
\usepackage{color}
\usepackage{booktabs}
\usepackage{nicematrix, tikz-cd, tikz}
\usetikzlibrary{matrix}
\usetikzlibrary{shapes,decorations,decorations.pathreplacing,calligraphy,arrows,calc,arrows.meta,fit,positioning}
\tikzset{
    -Latex,auto,node distance =1 cm and 1 cm,semithick,
    state/.style ={ellipse, draw, minimum width = 0.7 cm},
    point/.style = {circle, draw, inner sep=0.04cm,fill,node contents={}},
    bidirected/.style={Latex-Latex,dashed},
    el/.style = {inner sep=2pt, align=left, sloped}
}

\NiceMatrixOptions{transparent}
\usepackage{natbib}
\usepackage[colorlinks,citecolor=blue,urlcolor=blue,filecolor=blue]{hyperref}
\usepackage{graphicx}
\usepackage{capt-of}
\usepackage{bm}
\usepackage{algorithm}
\usepackage[noend]{algpseudocode}
\usepackage{float}
\usepackage[normalem]{ulem}
\usepackage{ifthen}
\usepackage{setspace}
\usepackage{multicol}

\newtheorem{thm}{Theorem}

\newtheorem{lemma}{Lemma}

\newcommand{\blind}{0}

\renewcommand{\arraystretch}{0.85}

\allowdisplaybreaks %

\begin{document}

\def\spacingset#1{
\renewcommand{\baselinestretch}%
{#1}\small\normalsize} \spacingset{1}

\if0\blind
{
  \title{\bf Comparison and Bayesian Estimation of Feature Allocations}
  \author{David B. Dahl \\
    \vspace{6pt}
    Department of Statistics, Brigham Young University \\
    Devin J. Johnson \\
    \vspace{6pt}
    Department of Statistical Science, Duke University \\
    R. Jacob Andros \\ 
    Department of Statistics, Brigham Young University}
  \maketitle
} \fi

\if1\blind
{
  \bigskip
  \bigskip
  \bigskip
  \begin{center}
    {\LARGE\bf Title}
\end{center}
  \medskip
} \fi

\bigskip

\begin{abstract}

Feature allocation models postulate a sampling distribution whose parameters
are derived from shared features.  Bayesian models place a prior distribution
on the feature allocation, and Markov chain Monte Carlo is typically used for
model fitting, which results in thousands of feature allocations sampled from
the posterior distribution.  Based on these samples, we propose a method to
provide a point estimate of a latent feature allocation. First, we introduce
FARO loss, a function between feature allocations which satisfies quasi-metric
properties and allows for comparing feature allocations with differing numbers
of features. The loss involves finding the optimal feature ordering among all
possible, but computational feasibility is achieved by framing this task as a
linear assignment problem. We also introduce the FANGS algorithm to obtain a
Bayes estimate by minimizing the Monte Carlo estimate of the posterior expected
FARO loss using the available samples. FANGS can produce an estimate other than
those visited in the Markov chain. We provide an investigation of existing
methods and our proposed methods.  Our loss function and search algorithm are
implemented in the \texttt{fangs} package in \texttt{R}.

\end{abstract}

\noindent%
\vfill

\newpage
\spacingset{1.5}

\section{Introduction}
\label{sec:intro}
Bayesian data analysis consists of two major tasks: sampling from the posterior
distribution (either directly or through some approximation), and then
summarizing those posterior draws in a way that is relevant to the problem at
hand. Parameter estimation is often a straightforward process, with the
posterior mean, median, mode, or a credible interval being sufficient in many
cases. However, for more complex parameters not contained in $\mathbb{R}^n$, a
different approach is necessary.

Feature allocation models from Bayesian nonparametrics are an example of such a
case where a more tailored estimation approach is needed. A feature allocation
$\rho = \{S_1,\ldots,S_K\}$ is a multiset of subsets (i.e., features) such that
each subset $S_k \subseteq \{1,\ldots,n\}$ is nonempty. Being a multiset, the
order of the subsets is irrelevant, and there may be duplicates. In contrast to
partitions, these subsets are not necessarily mutually exclusive nor exhaustive.
For some subset $S_k \in \rho$, we say that item $i$ has feature $k$ if $i \in
S_k$, and items $i$ and $j$ share feature $k$ if $i \in S_k$ and $j \in S_k$. In
model construction, features are often used to arrange data $y_1,\ldots,y_n$
such that data sharing a feature $k$ have some aspect of the data model in
common. A feature allocation $\rho$ can alternatively be represented by a binary
matrix $Z$, with $n$ rows and $K$ columns, containing $1$ in the $(i,j)$ element
if and only if $i \in S_k$, for $k=1,\ldots,K$.  \citet{griffithsIBP} defined an
equivalence class for these binary feature allocation matrices based on a
left-ordering function $lof(\cdot)$, which reorders the columns of a feature
allocation matrix from left to right by the binary number of each column in
descending order. Under this equivalence class, the order of subsets in $\rho$
is irrelevant; likewise, the order of the columns (or features) in $Z$ is not
important. Furthermore, under the left-ordering function, a column of all zeros
represents a null column (or an empty feature), so adding a column of zeros to
$Z$ does not change its equivalence.

Feature allocations have been used for a wide range of applications, most
frequently in biostatistics. Such applications include disease classification
\citep{warr2021attraction}, cell subpopulation identification \citep{lui2021},
tumor heterogeneity inference \citep{muller2015}, and symptom-disease
relationship modeling \citep{ni2020bayesian}. These same studies have also
explored methods for sampling feature allocations using MCMC. But again, while
obtaining posterior samples is crucial, it is only part of the inference
problem.  We propose the FARO loss function and FANGS search algorithm that work
in tandem to provide a feature allocation point estimate using the posterior
samples. FARO loss is not restricted to comparing matrices with a common number
of features and does not discriminate based on feature ordering. The proposed
FANGS search algorithm can then quickly and repeatedly compute Monte Carlo
estimates of the expected FARO loss for different candidate estimates. It uses
these candidates to optimize the feature allocation space and will almost
certainly find a better feature allocation than those visited by the Markov
chain. Our loss function and search algorithm are implemented in the
\texttt{fangs} package for \texttt{R}.

The paper proceeds as follows. In Section \ref{sec:existing_work}, we provide
additional background on feature allocations and review the existing literature.
In Section \ref{sec:new_loss}, we detail a new loss function and the unique
advantages that it holds over the previous losses described in Section
\ref{sec:existing_work}. In Section \ref{sec:fangs}, we explain the intuition
behind our FANGS search algorithm that works in tandem with the FARO loss
function proposed in Section \ref{sec:new_loss}. Section \ref{sec:verifications}
contains a comparison through simulation studies of the  existing search
algorithms with our proposed FANGS method. Finally, Section \ref{sec:conclusion}
provides a summary of the contributions these methods make to the literature on
Bayesian feature allocation models.

\section{Existing Literature on Feature Allocations}
\label{sec:existing_work}
In this section, we review existing loss function for feature allocations and
existing search algorithms for feature allocation estimation.  This provides a
context for our proposed FARO loss function (in Section \ref{sec:new_loss}) and
FANGS search algorithm (in Section \ref{sec:fangs}).

\subsection{Background on Loss Functions}

In Bayesian statistics, we typically choose a loss function that can measure the
distance between two objects in the parameter space (e.g., an estimator
$\hat{\theta}$ and the true parameter value $\theta$). A higher loss indicates
more dissimilarity between the estimator and the parameter value. Then, the
posterior expectation of that loss function is minimized to find the Bayesian
estimator of the parameter. In the case of feature allocations, for some loss
function $L(\rho, \hat\rho)$ used to measure the loss in estimating $\rho$ with
$\hat\rho$, the Bayes estimate $\hat\rho^*$ is:
\vspace{-0.45cm}
\begin{equation}
\label{eq_generic_loss}
\hat\rho^* = \underset{\hat\rho}{\text{argmin}}\hspace{3pt}\mathbb{E}(L(\rho, \hat\rho) \mid \mathcal{D})
\quad \quad \text{or} \quad \quad
\hat{Z}^* = \underset{\hat{Z}}{\text{argmin}}\hspace{3pt}\mathbb{E}(L(Z, \hat{Z}) \mid \mathcal{D}),
\vspace{-0.55cm}
\end{equation}
where $\mathcal{D}$ is the observed data, and $\hat{Z}^*$ and $L(Z, \hat{Z})$
are parameterized in binary matrix notation and are equivalent to $\hat\rho^*$
and $L(\rho, \hat\rho)$. Without loss of generality, we assume that if the
estimator (i.e., $\hat\rho$ or $\hat{Z}$) is equal to the true parameter (i.e.,
$\rho$ or $Z$), then the loss function evaluates to zero. Otherwise, the loss
evaluates to a positive value that represents the ``cost'' associated with using
the chosen estimator instead of the truth. In most applications, the posterior
expectation in (\ref{eq_generic_loss}) is approximated using posterior samples:
\vspace{-0.4cm}
\begin{equation}
    \mathbb{E}(L(\rho, \hat\rho) \mid \mathcal{D}) \approx \frac{1}{B} \sum_{b=1}^B L(\rho_b, \hat\rho)
    \quad \quad \text{or} \quad \quad
    \mathbb{E}(L(Z, \hat{Z}) \mid \mathcal{D}) \approx \frac{1}{B} \sum_{b=1}^B L(Z_b, \hat{Z}),
    \label{eq:estExpLoss}
    \vspace{-0.1cm}
\end{equation}
where $\rho_1,\ldots,\rho_B$ or $Z_1,\ldots,Z_B$ are $B$ feature allocation
samples from a posterior distribution $p(\rho \, | \, \mathcal{D})$ or $p(Z \, |
\, \mathcal{D}$). Samples are often obtained through considerable effort from
several MCMC chains; we simply assume they are available. We seek to find an
estimate using the available samples that summarizes the feature allocation
distribution.

\subsection{Existing Loss Functions for Feature Allocations}

Loss functions for estimating parameters contained in $\mathbb{R}^n$ --- such as
squared error loss or absolute error loss --- are well studied and frequently
applied.  In this subsection, we review the relatively modest literature on loss
functions for feature allocations.

\subsubsection{Zero-One Loss}

\citet{gershman} proposed using the maximum a posteriori (MAP) feature
allocation in the context of the distance dependent Indian Buffet Process
(ddIBP). \citet{warr2021attraction} successfully implemented this method in a
classification study of Alzheimer's disease neuroimaging. \citet{muller2015}
devised an objective function, similar to a penalized likelihood function, to
find the MAP. Recall that the MAP estimator, in theory, relies on the zero-one
loss function. When using the MAP estimator in feature allocation summarization,
we are required to have either: i) access to a computable posterior density that
can be optimized, or ii) enough posterior samples and small enough matrix
dimensions that the same feature allocation $Z$ has been visited multiple times
in the Markov chain.  The first requirement may be tedious since a set of
posterior samples would not be enough to carry out estimation, and the second is
unlikely to be met in nontrivial applications. Finally, the MAP estimate may not
well represent the ``center'' of the posterior feature allocation distribution,
and could be problematic in multimodal or heavily skewed distributions.

\subsubsection{Loss Based on Pairwise Similarity}

\citet{lui2021} modeled cytometry data using a latent feature allocation model
and based their loss function on the pairwise similarity matrix (PSM), denoted
by $\Psi$. The $(i,j)$ element of $\Psi$ gives the expected number of features
shared by items $i$ and $j$.  The PSM $\Psi$ is estimated from posterior samples
$Z_1, \ldots, Z_B$ as $ \hat\Psi_{ij} = \frac{1}{B}\sum_{b=1}^B Z_b Z_b' $,
where $Z'$ is the transpose of $Z$ and the summation is elementwise.
\citet{lui2021} proposed a feature allocation estimate $\hat Z$ which minimizes
the sum of squared distances from each element of $\Psi$:

\vspace{-3pt}
\begin{equation}
    \hat{Z} = \underset{Z}{\text{argmin}} \sum_{i=1}^n \sum_{j=1}^n ((Z Z')_{ij} - \Psi_{ij})^2.
\end{equation}
\vspace{-23pt}

\begin{figure}[tb]
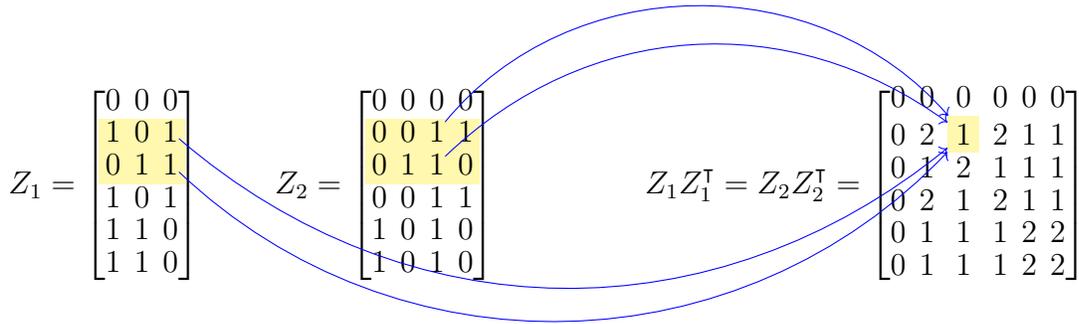

    \begin{center}
        $ Z_1 =
        \setlength\arraycolsep{2.5pt}
        \begin{bmatrix}[name=G]
            0 & 0 & 0 \\
            \rowcolor{yellow!40}
            1 & 0 & 1 \\
            \rowcolor{yellow!40}
            0 & 1 & 1 \\
            1 & 0 & 1 \\
            1 & 1 & 0 \\
            1 & 1 & 0
        \end{bmatrix} \hspace{1cm}
        Z_2 =
        \begin{bmatrix}[name=H]
            0 & 0 & 0 & 0 \\
            \rowcolor{yellow!40}
            0 & 0 & 1 & 1 \\
            \rowcolor{yellow!40}
            0 & 1 & 1 & 0 \\
            0 & 0 & 1 & 1 \\
            1 & 0 & 1 & 0 \\
            1 & 0 & 1 & 0
        \end{bmatrix}
        \hspace{2cm} 
        Z_1 Z_1^\intercal = Z_2 Z_2^\intercal = 
        \setlength\arraycolsep{2.5pt}
        \begin{bmatrix}[name=J]
            0 & 0 & 0 & 0 & 0 & 0 \\
            0 & 2 & \colorbox{yellow!40}{$1$} & 2 & 1 & 1 \\
            0 & 1 & 2 & 1 & 1 & 1 \\
            0 & 2 & 1 & 2 & 1 & 1 \\
            0 & 1 & 1 & 1 & 2 & 2 \\
            0 & 1 & 1 & 1 & 2 & 2 
        \end{bmatrix}$ \hspace{1.5cm}
        \tikz [remember picture, overlay] \draw [blue,->] (G-2-3) to [bend right=40] (J-2-3) ;
        \tikz [remember picture, overlay] \draw [blue,->] (G-3-3) to [bend right=45] (J-2-3) ;
        \tikz [remember picture, overlay] \draw [blue,->] (H-2-3) to [bend left=50] (J-2-3) ;
        \tikz [remember picture, overlay] \draw [blue,->] (H-3-3) to [bend left=40] (J-2-3) ;
        
    \end{center}
    \caption{The distinct feature allocations $Z_1$ and $Z_2$ map to the same adjacency matrix.}
    \label{fig:ZmatVsAdjMat}
\end{figure}

This approach mirrors the literature of partition estimation from
\citet{dahl.2006} in which the sum of squares to a clustering adjacency matrix
was minimized. \citet{lau.green.2007} proposed minimizing the posterior
expectation of the popular Binder loss. Later, \citet{dahl.newton.2007} showed
that the least-squares estimate of a partition is actually equivalent to
minimizing Binder loss.  A key to this equivalence relies on the fact that a
partition is unique to its adjacency matrix. Unfortunately, a feature allocation
$Z$ is not always unique to its adjacency matrix $Z Z'$, as demonstrated in
Figure \ref{fig:ZmatVsAdjMat}. Two distinct feature allocations are shown here,
one with three features and one with four features. Upon inspection, these two
feature allocations seem to be quite distinct, and yet they still map to the
same adjacency matrix. Hence, a loss function using the adjacency matrix would
not distinguish between these two feature allocations. This problem is amplified
as $n$ and $K$ increase.

\subsubsection{Hamming Distance}

\citet{ni2020bayesian} proposed a double feature allocation (DFA) model for
patient-disease data. They used a loss function conditioned on $\hat{K}$, the
modal number of features from the samples. The distance between a given binary
matrix $Z \in \{0,1\}^{n \times \hat{K}}$ and an estimate $\hat{Z}$ was defined
as the minimum Hamming distance between $Z$ and the set of all column
permutations of $\hat{Z}$, which they denote using $\pi(\hat{Z})$ (Equation
\ref{eq:hamming}). The Hamming distance between two matrices $Z$ and $\hat{Z}$
(with the same number of columns) simply counts the number of $(i,j)$ indices
that disagree. Note that $\mathds{I}(Z_{ij} \neq \hat{Z}_{ij} )$ is 1 if the
$(i,j)$ entries of $Z$ and $\hat{Z}$ disagree, and is 0 if they are the same.
\vspace{-0.4cm}
\begin{equation}
    \mathcal{H}(Z, \hat{Z}) = \sum_{i=1}^n \sum_{j=1}^K \mathds{I}(Z_{ij} \neq \hat{Z}_{ij} ) \hspace{1.5cm} L(Z,\hat{Z}) = \underset{\pi}{\text{min}} \hspace{2pt} \mathcal{H}(Z, \pi(\hat{Z})) 
    \label{eq:hamming}
    \vspace{-0.4cm}
\end{equation}
However, their loss function forces the number of features in their estimated
feature allocation to match the expected number of features. In practice, any
posterior samples whose number of features differ from the observed modal number
of features ($\hat{K}$) are discarded when estimating the feature allocation. In
addition, examining all column permutations of a feature allocation matrix gets
extremely difficult for high values of $K$, with order $\mathcal{O}(K!)$. For
example, $10!$ is approximately 3.6 million, and examining that many different
permutations takes over 6 seconds (see Table \ref{tab:permVsFaro} in Section
\ref{subsec:assignment}). To find the feature allocation with the lowest
expected loss among the $B$ posterior samples, the loss would have to be
computed $B^2$ times, so spending even a few seconds on each column alignment is
not viable.

In spite of the computational disadvantages, column-permuted Hamming distance
may be the most intuitive of all the losses discussed so far. Binder loss is
commonly used for partitions \citep{dahl2021search} and can be expressed in
terms of Hamming distance \citep{wade.ghah.2018}. It would then be logical to
use the Hamming distance to formulate a loss function for feature allocation
estimation, which we explore further in Section 3.

\subsection{Existing Search Algorithms for Feature Allocations}

As was the case with loss functions for feature allocations, there is only a
relatively modest literature on search algorithms for minimizing the posterior
expected loss in feature allocation estimation.  A loss function without an
effective means to minimize its posterior expectation is of little practical use
when seeking to find a Bayes estimator.

The studies discussed in the previous section included the simulations from the
AIBD paper \citep{warr2021attraction}, the cytometry data model \citep{lui2021},
and the DFA model for patient-disease relationships \citep{ni2020bayesian}. The
loss functions in these studies were different; the first used zero-one loss,
the second used a sum of squares function on the PSM, and the last used Hamming
distance. However, they all used the same search algorithm, namely, the ``draws
method'' \citep{dahl.2006}. The draws method comprises finding the feature
allocation among those sampled that minimizes the estimated expected loss:

\begin{equation}
    \label{eq:draws_method}
    \hat{Z}^* = \underset{\hat{Z}}{\text{argmin}} \hspace{2pt} \frac{1}{B} \sum_{b=1}^B L(Z_b, \hat{Z}) \hspace{2pt} \text{ for } \hat{Z} \in \{ Z_b: b=1, \hdots, B \}
\end{equation}
In the case of the DFA model, the draws method is restricted to only search
among those sampled feature allocations whose number of features equals the
modal number of features $\hat{K}$. That is, rather than selecting from all
posterior samples $\hat{Z} \in \{ Z_b: b=1, \hdots, B \}$, the algorithm is
limited to choosing from $\hat{Z} \in \{ Z^{n \times K}: K = \hat{K} \}
\subseteq \{ Z_b: b=1, \hdots, B \} $.

Though computationally efficient and widely applicable, the draws method is
clearly flawed --- its chosen estimate is restricted to the samples visited in
the Markov chain \citep{dahl2021search}. There almost certainly exists an
estimate within the parameter space (but not sampled in the chains) that yields
lower loss than the minimizing draw. At the same time, it would be impractical
to conduct an exhaustive search over all possible feature allocation matrices;
for a given binary $n \times K$ matrix, there exist $2^{n\times K}$ possible
objects. Even if $n=25$ and $K=4$, this would entail searching through more than
$10^{29}$ matrices.

Besides the draws method, we are aware of one other search algorithm used by
\citet{zeng2018} in a tumor subclone identification study. They called their
method SIFA, short for subclone identification using feature allocations. In the
SIFA example, all posterior samples had the same number of features. They
sequentially reordered the columns of each sample by finding the column
permutation that minimized the Hamming distance between $Z_i$ and $Z_{i-1}$ for
$i = 2,\ldots,B$. Within these aligned feature allocations, the element-wise
mode was computed for all $(i,j)$ entries to form the feature allocation
estimate. Their approach computes all possible column reorderings, much like
\citet{ni2020bayesian}, and thus becomes infeasible for increasing $K$. This
idea is intriguing as it is not stochastic and can explore more than just the
sampled draws, but it comes at a heavy computational expense.

In short, the ideal search algorithm should be computationally efficient for
increasing $K$, more expansive than the draws method, and avoid the cost of an
exhaustive search.

\section{Feature Allocations Reordered Optimally (FARO)}
\label{sec:new_loss}
In this section we introduce FARO (an acronym for \textbf{f}eature
\textbf{a}llocations \textbf{r}eordered \textbf{o}ptimally) loss as a loss
function between two feature allocations. As we explain in this section, our
FARO loss is simply the minimum generalized Hamming distance between all column
permutations of two feature allocation matrices. Importantly, FARO  loss can
compare matrices having different numbers of features $K$. Minimization problems
based on computing the expected FARO loss only need access to a set of sampled
feature allocations; a closed-form posterior density is not necessary. FARO loss
is also more computationally efficient than previous loss functions and can
therefore be evaluated repeatedly and rapidly. This will be crucial for the
feasibility of our FANGS search algorithm that we introduce in Section
\ref{sec:fangs}.

\subsection{Relation to Hamming Distance}
\label{subsec:hamming_dist}

\citet{binder.1978} loss is a popular loss function for partitions that can be
expressed in terms of Hamming distances \citep{wade.ghah.2018} between adjacency
matrices from partitions. Our FARO loss for feature allocations is also based on
Hamming distance, but the Hamming distances are now computed among the binary
matrices (e.g., $Z$) instead of adjacency matrices (e.g., $Z Z'$). Using Hamming
distance to compare feature allocations is a promising start and is an important
component of FARO loss inspired by the existing literature, but FARO loss makes
four key additional contributions. First, FARO loss can compare feature
allocations with different numbers of features, thus permitting all posterior
samples to aid in estimation (Section \ref{subsec:imb_dimensions}). Second, we
introduce the use of unequal penalties when computing Hamming distance between
feature allocations (Section \ref{subsec:diff_penalties_hamming}), following the
flexibility of \citet{binder.1978} loss and generalized variation of information
\citep{dahl2021search} in the partition literature. Our generalization with
differential costs induces control over the sparsity when estimating a feature
allocation. Third, we note that finding the minimum Hamming distance for all
column permutations does not require an exhaustive search. Rather, we formulate
this as a linear assignment problem (Section \ref{subsec:assignment}) which is
well-studied in the computer science literature.  This formulation allows for
rapid computation of FARO loss, opening the possibility of repeated computation
of the posterior expected loss in our proposed search algorithm. Fourth, we
prove that FARO loss is a quasi-metric in Section \ref{subsec:metric}.

\subsection{Imbalanced Dimensions}
\label{subsec:imb_dimensions}

The Hamming distance between $Z$ and $\hat{Z}$ is a weighted sum of
disagreements between their corresponding indices and is only defined for
matrices with the same dimensions. \citet{ni2020bayesian} implicitly assign
infinite loss for any feature allocation whose number of features $K$ does not
equal the modal number of features $\hat{K}$. Thus any posterior samples with $K
\neq \hat{K}$ are completely discarded in their estimation of the feature
allocation. In another paper, \citet{ni-cmc} proposed a consensus Monte Carlo
(CMC) approach for scaling feature allocations modeling big data where they
remove the excess columns from the matrix with more features.  In
\cite{zeng2018}, the sampling process was such that all feature allocation draws
had the exact same number of features, eliminating the need to address the
dimensionality problem. Our approach is different from all three of these
approaches. FARO loss accounts for an imbalanced number of columns by augmenting
the matrix having fewer columns with empty features (i.e., columns of zeros).
This does not alter equivalence under the left-ordering equivalence class
\citep{griffithsIBP} and simplifies the comparison between matrices based on
Hamming distance.

\subsection{Generalized Hamming Distance}
\label{subsec:diff_penalties_hamming}

In \citet{ni2020bayesian}, \citet{zeng2018}, and in our own discussion to this
point, Hamming distance $\mathcal{H}(Z, \hat{Z})$ has been defined as the
summation of all disagreements between indices of two feature allocation
matrices.
\begin{equation}
    \mathcal{H}(Z, \hat{Z}) = \sum_{i=1}^n \sum_{j=1}^K \mathds{I}(Z_{ij} \neq \hat{Z}_{ij}) = \sum_{i=1}^n \sum_{j=1}^K \big[ \mathds{I}(Z_{ij} = 1 \cap \hat{Z}_{ij} = 0) + \mathds{I}(Z_{ij} = 0 \cap \hat{Z}_{ij} = 1) \big] \notag
\end{equation}
Implicitly, there is a penalty of 1 when $Z_{ij} \neq \hat{Z}_{ij}$ regardless
of whether this disagreement arises from $Z_{ij} = 1$ and $\hat{Z}_{ij}=0$ or
from $Z_{ij} = 0$ and $\hat{Z}_{ij}=1$.  \citet{binder.1978} proposed a
partition loss with two distinct penalty parameters $a$ and $b$.  That is, in
its equal-cost form, we would have $a=b=1$. \citet{dahl2021search} showed that
differential costs $a \not= b$ allows for tuning of the number of clusters
present in a partition estimate. In the same way, we propose that the penalties
$a$ and $b$ can be used for feature allocation estimation.  Without loss of
generality, let $a,b \in (0,2)$ and $a+b=2$.  We call this function the
generalized Hamming distance\footnote{This should not be confused with the
definition of generalized Hamming distance given by \citet{bookstein2002} in the
computer science literature to compare bitmaps and bitstrings.} $\mathcal{H}_G
(Z, \hat{Z})$:
\begin{equation}
    \mathcal{H}_G (Z, \hat{Z}) = \sum_{i=1}^n \sum_{j=1}^K \big[ a \cdot \mathds{I}(Z_{ij} = 1 \cap \hat{Z}_{ij} = 0) + b \cdot \mathds{I}(Z_{ij} = 0 \cap \hat{Z}_{ij} = 1) \big]
    \label{eq:general_hamming}
\end{equation}

Our FARO loss is simply the minimum generalized Hamming distance between all
column permutations of two feature allocation matrices. As is the case for
controlling the number of clusters in partition estimation
\citep{dahl2021search}, the ability to adjust the penalties controls the
sparsity in feature allocation estimation. This becomes particularly important
when dealing with samples having extremely high $K$, as they tend to yield less
interpretable feature allocation estimates. For the remainder of the paper, we
refer to generalized Hamming distance instead of the original Hamming distance.

\subsection{The Linear Assignment Problem}
\label{subsec:assignment}

Even after augmenting a smaller matrix with columns of zeros, the generalized
Hamming distance between two matrices is dependent on the column ordering.
Recall that the FARO loss is the minimum generalized Hamming distance between
all column permutations of two feature allocation matrices. Under the
left-ordering equivalence class, the ordering of the columns should not matter
--- any reordering of the columns of a binary feature allocation matrix are
equivalent. As in \citet{ni2020bayesian}, we would like to remove this
dependency by optimally reordering or aligning the columns (features) of two
binary matrices $Z, \hat{Z}$ such that the distance is minimized. Their method
computes and compares all possible column orderings of each $Z$ in order to find
this minimizing alignment, which was feasible in their study because the
matrices of interest usually contained six or fewer features. However, for a
case where $K=10$, there would be $10! \approx 3.63 \times 10^6$ possible
alignments (see Table \ref{tab:permVsFaro}). Clearly, carrying out this many
evaluations of the distance between a candidate and many posterior samples would
be computationally prohibitive.

\begin{figure}[tb]
\begin{centering}
    \begin{tikzpicture}[
            agents/.style={circle, draw=blue!60, fill=blue!5, very thick, minimum size=9mm},
            tasks/.style={circle, draw=gray!90, fill=gray!10, very thick, minimum size=9mm},
            ]
        \node[] (atext) at (0,0) {\large Agents};
        \node[agents] (a1) [below=.2cm of atext] {\textbf{1}};
        \node[agents] (a2) [below=.2cm of a1] {\textbf{2}};
        \node[agents] (a3) [below=.2cm of a2] {\textbf{3}};
        \node[agents] (a4) [below=.2cm of a3] {\textbf{4}};
        \node[tasks] (t1) [right=1.6cm of a1] {\textbf{1}};
        \node[tasks] (t2) [right=1.6cm of a2] {\textbf{2}};
        \node[tasks] (t3) [right=1.6cm of a3] {\textbf{3}};
        \node[tasks] (t4) [right=1.6cm of a4] {\textbf{4}};
        \node[] (ttext) [above=.3cm of t1] {\large Tasks};
        \node[] (align1) at (1.2,-5.4) {Alignment 1: $\mathcal{H}_G=23$};

        \path (a1) edge node[el,above,pos=.2] {$\mathbf{3}$} (t2);
        \path (a2) edge node[el,below,pos=.2] {$\mathbf{5}$} (t1);
        \path (a3) edge node[el,above,pos=.2] {$\mathbf{10}$} (t4);
        \path (a4) edge node[el,below,pos=.2] {$\mathbf{5}$} (t3);

        \node[] (atextx) [right=1cm of ttext] {\large Agents};
        \node[agents] (a1x) [below=.2cm of atextx] {\textbf{1}};
        \node[agents] (a2x) [below=.2cm of a1x] {\textbf{2}};
        \node[agents] (a3x) [below=.2cm of a2x] {\textbf{3}};
        \node[agents] (a4x) [below=.2cm of a3x] {\textbf{4}};
        \node[tasks] (t1x) [right=1.6cm of a1x] {\textbf{1}};
        \node[tasks] (t2x) [right=1.6cm of a2x] {\textbf{2}};
        \node[tasks] (t3x) [right=1.6cm of a3x] {\textbf{3}};
        \node[tasks] (t4x) [right=1.6cm of a4x] {\textbf{4}};
        \node[] (ttextx) [above=.3cm of t1x] {\large Tasks};
        \node[] (align2) [right=1.15cm of align1] {Alignment 2: $\mathcal{H}_G=17$};

        \path (a1x) edge node[el,above,pos=.15] {$\mathbf{5}$} (t3x);
        \path (a2x) edge node[el,below,pos=.85] {$\mathbf{6}$} (t4x);
        \path (a3x) edge node[el,above,pos=.85] {$\mathbf{4}$} (t1x);
        \path (a4x) edge node[el,below,pos=.15] {$\mathbf{2}$} (t2x);

        \node[] (atexty) [right=1 cm of ttextx] {\large Agents};
        \node[agents] (a1y) [below=.2cm of atexty] {\textbf{1}};
        \node[agents] (a2y) [below=.2cm of a1y] {\textbf{2}};
        \node[agents] (a3y) [below=.2cm of a2y] {\textbf{3}};
        \node[agents] (a4y) [below=.2cm of a3y] {\textbf{4}};
        \node[tasks] (t1y) [right=1.6cm of a1y] {\textbf{1}};
        \node[tasks] (t2y) [right=1.6cm of a2y] {\textbf{2}};
        \node[tasks] (t3y) [right=1.6cm of a3y] {\textbf{3}};
        \node[tasks] (t4y) [right=1.6cm of a4y] {\textbf{4}};
        \node[] (ttexty) [above=.3cm of t1y] {\large Tasks};
        \node[] (align3) [right=1.15cm of align2] {Alignment 3: $\mathcal{H}_G=11$};

        \path (a1y) edge node[el,above,pos=.3] {$\mathbf{3}$} (t2y);
        \path (a2y) edge node[el,above,pos=.65] {$\mathbf{2}$} (t3y);
        \path (a3y) edge node[el,above,pos=.5] {$\mathbf{4}$} (t1y);
        \path (a4y) edge node[el,below,pos=.4] {$\mathbf{2}$} (t4y);
        
    \end{tikzpicture}
    \caption{An example of the linear assignment problem in the context of
    assigning an agent (a column of $Z$) to a task (a column of $\hat{Z}$).
    While the first two alignments (left and middle) lead to costs of 23 and 17,
    the last alignment yields the optimal cost of 11.} \label{fig:alignments}
\end{centering}
\end{figure}
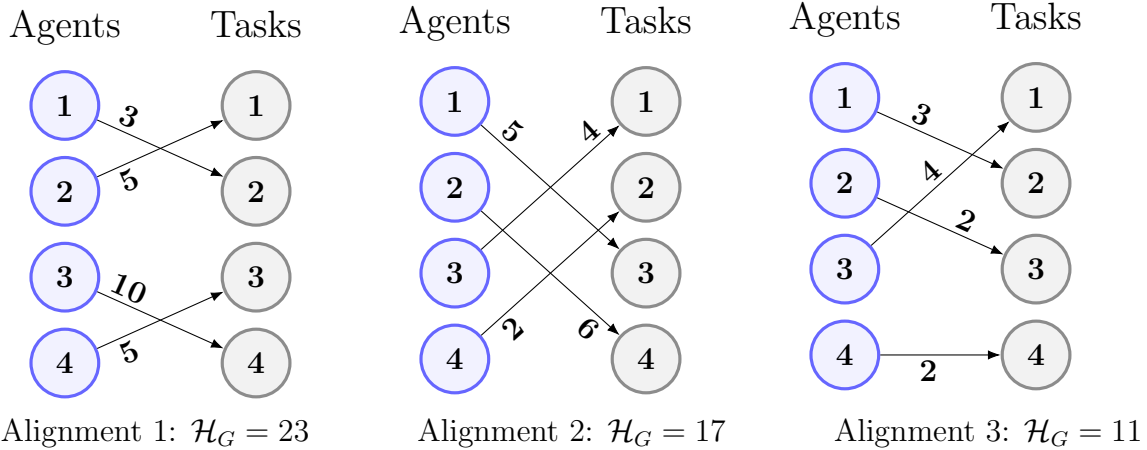

One of the key contributions of this paper is recognizing that the optimal
alignment task is a linear assignment problem, for which fast algorithms are
available.  Rapid computation of the optimal alignment permits repeated
computation of FARO loss, which enables our FANGS method (Section
\ref{sec:fangs}) to be feasible for feature allocation estimation. The linear
assignment problem (in its balanced form) is often described in the context of
agents and tasks; each of $K$ agents can be assigned to one and only one of $K$
tasks. A small example is illustrated in Figure \ref{fig:alignments} and
Equation \ref{eq:costmatrix}. In our case, the ``agents'' are the $K$ features
from $Z$ and the ``tasks'' are the $K$ features from $\hat{Z}$. There is a cost
for assigning a particular agent to a particular task; in our case, this
represents the fact that there is some generalized Hamming distance between a
particular column of $Z$ and a particular column of $\hat{Z}$. The end goal is
to assign each agent to a task such that the sum of the costs is minimized,
while avoiding an exhaustive search of all possible alignments.

The linear assignment problem avoids an exhaustive search of all alignments
because it is essentially a constrained minimization problem. The objective
function is the trace of a transposed permutation matrix $\mathbf{X}$ multiplied
by a cost matrix $\mathbf{C}(Z, \hat{Z})$. The $ K \times K $ cost matrix is
constructed such that $ \mathbf{C}_{ij}(Z,\hat{Z}) $ represents the cost (or
generalized Hamming distance) of lining up feature $ i $ of $Z$ with feature $ j
$ of $\hat{Z}$. The constraints of the minimization problem follow directly from
the definition of the permutation matrix: $\mathbf{X}$ is a $K \times K$ binary
matrix having exactly one nonzero entry in each row and column
\citep{brualdi2006}. In other words, we must choose exactly one item in each row
and column in order to define a valid alignment of the matrices such that the
total cost (total generalized Hamming distance) is minimized. In the simple
example in Figure \ref{fig:alignments} and Equation \ref{eq:costmatrix}, the
minimizing alignment is $x_{12} = x_{23} = x_{31} = x_{44} = 1$, leading to a
cost of $3+2+4+2=11$. For our purposes, this implies that column 1 of $Z$ should
be aligned with column 2 of $\hat{Z}$, column 2 of $Z$ with column 3 of
$\hat{Z}$, and so forth, such that the FARO loss is 11.

\def\SmallColSep{\setlength{\arraycolsep}{1pt}}
\begin{align}
    &\mathbf{X} =
    \begingroup\SmallColSep
    \begin{bmatrix}
        x_{11} & \colorbox{yellow!40}{$x_{12}$} & x_{13} & x_{14} \\[-0.25em]
        x_{21} & x_{22} & \colorbox{yellow!40}{$x_{23}$} & x_{24} \\[-0.25em]
        \colorbox{yellow!40}{$x_{31}$} & x_{32} & x_{33} & x_{34} \\[-0.25em]
        x_{41} & x_{42} & x_{43} & \colorbox{yellow!40}{$x_{44}$}
    \end{bmatrix}
    \endgroup
    \hspace{9pt}
    \mathbf{C}(Z, \hat{Z}) = 
    \begin{bmatrix}
        14 & \colorbox{yellow!40}{3}  & 5 & 8 \\[-0.25em]
        5  & 12 & \colorbox{yellow!40}{2} & 6 \\[-0.25em]
        \colorbox{yellow!40}{4}  & 7  & 7 & 10 \\[-0.25em]
        9  & 2  & 5 & \colorbox{yellow!40}{2}
    \end{bmatrix}
    \hspace{0.8cm}
    \begin{aligned}
    &\textbf{Objective Function: } \\ 
    & f(Z,\hat{Z}) = \text{tr}\big(\mathbf{X}^\prime \, \mathbf{C}(Z,\hat{Z}) \big)
    \end{aligned}
    \notag \\
    &\textbf{Constraints: } \displaystyle \sum_{i=1}^{4} x_{ij} = \displaystyle \sum_{j=1}^{4} x_{ij} = 1 \, \text{ and } \, x_{ij} \in \{0, 1\}; \hspace{3pt} i,j \in \{1,\ldots,4\}
    \label{eq:costmatrix}  
\end{align}
\vspace{-0.7cm}

\begin{table}[tb]
    \centering
    \begin{tabular}{rrr@{\hspace{2pt}}rrrr@{\hspace{2pt}}}
        \toprule[1.5pt]
        & & \multicolumn{2}{c}{\textbf{Traditional Approach}} & & \multicolumn{2}{c}{\textbf{Our Approach}}  \\
        & & \multicolumn{2}{c}{\textbf{(All Permutations)}} & & \multicolumn{2}{c}{\textbf{(Linear Assignment)}}  \\
        \cmidrule{3-4} \cmidrule{6-7} 
        $K$ & & $K!$ & Time (ms) & & $K^3$ & Time (ms) \\
        \midrule
        4 & & 24 & 0.04 & & 64 & 0.02 \\
        6 & & 720 & 0.83 & & 216 & 0.02 \\
        8 & & 40,320 & 58.12 & & 512 & 0.03 \\
        10 & & 3,628,800 & 6,444.61 & & 1,000 & 0.05 \\
        \bottomrule[1.5pt]
    \end{tabular}
    \bigskip \caption{Mean time (over 100 replications, in milliseconds) to find
    the minimum Hamming distance between two randomly sampled matrices with
    $n=100$ rows and $K \in \{4,6,8,10\}$ columns using (left) all possible
    permutations and (right) the J\"onker-Volgenant algorithm. Note that it took
    more than 6 seconds to examine all permutations for $K=10$, whereas the
    J\"onker-Volgenant algorithm only took a fraction of a millisecond.}
    \label{tab:permVsFaro}
\end{table}

The linear assignment problem is quickly solved by the
\citet{jonker1987shortest} algorithm in the computer science literature, which
builds on an earlier solution called the Hungarian algorithm. Their algorithm is
guaranteed to return an optimal alignment and does so much faster than an
exhaustive search. For our software implementation of the FARO loss, we use an
implementation of the J\"onker-Volgenant algorithm provided in the Rust crate
\texttt{lapjv} \citep{pkg.lapjv}, which has $\mathcal{O}(K^3)$ complexity.
Recall that the distance computation for feature allocations from
\citet{ni2020bayesian} examined all column permutations of each matrix making it
viable only for small values of $K$ because it has order $\mathcal{O}(K!)$. A
rapidly computed loss function, especially when $K$ is large, is critical in
order to successfully implement a more thorough search algorithm. Table
\ref{tab:permVsFaro} shows the mean time to compute the minimum Hamming distance
for both approaches. Spending even 6 seconds to compute the loss between two
feature allocations with $K=10$ features using an exhaustive search may be too
much for a robust optimization procedure, but the J\"onker-Volgenant algorithm
makes this trivial even for large $K$.

\subsection{Quasi-Metric Properties of FARO Loss}
\label{subsec:metric}

A metric satisfies three properties: 1) the identity of indiscernibles, 2) symmetry, and 3) the triangle inequality. A quasi-metric is a function that satisfies the identity of indiscernibles and triangle inequality but does not need to satisfy the symmetry property.  In this section we prove that FARO loss is a quasi-metric when $a \not= b$ and a metric when $a=b=1$.
We also discuss why each property is desirable for a feature allocation loss function.

\begin{thm}\label{faro_metric}
	FARO loss is a quasi-metric on the feature allocation space.
\end{thm}

Before proceeding with the proof, it is helpful to point out that Hamming distance (under the equal penalty assumption) has already been proven to be a metric by \citet{robinson2003}. It follows naturally that generalized Hamming distance satisfies the properties of a quasi-metric, which we prove in the appendix. Given that the generalized Hamming distance is a quasi-metric, we seek to prove that the minimum of $m = K!$ generalized Hamming distances is also a quasi-metric (where $K$ is the number of columns in each feature allocation matrix after augmentation). To do this, we denote the FARO loss between two binary feature allocation matrices $Z, \hat{Z}$ as follows, where each $L_i$ is simply the generalized Hamming distance between $Z$ and a distinct permutation of the columns of $\hat{Z}$. 
\[ L(Z, \hat{Z}) = \text{min} \{ L_1(Z, \hat{Z}), L_2(Z, \hat{Z}), \ldots, L_m(Z, \hat{Z}) \}. \]
A loss function $L$ satisfies the identity of indiscernibles when $L(Z, \hat{Z}) = 0$ if and only if $Z = \hat{Z}$. First, suppose $L(Z, \hat{Z}) = 0$. Then there exists a loss function $L_i$ among the $m$ quasi-metrics that satisfies $L_i(Z, \hat{Z}) = 0$. Since each $L_i$ is a quasi-metric and satisfies the identity of indiscernibles, it follows that $Z = \hat{Z}$. Conversely, suppose $Z = \hat{Z}$. Then for at least one quasi-metric $L_i \in \{L_1, L_2, \ldots, L_m\}$, we have $L_i(Z, \hat{Z}) = 0$. Because at least one $L_i$ is equal to zero, we have that $\min \{ L_1(Z, \hat{Z}), L_2(Z, \hat{Z}), \ldots, L_m(Z, \hat{Z}) \} = 0$ so $L(Z, \hat{Z}) = 0$. Therefore $L(Z, \hat{Z}) = 0 \iff Z = \hat{Z}$, so FARO loss satisfies the identity of indiscernibles. A loss function with this property conforms closely to our intuition of distance, since two distinct feature allocations should have nonzero distance between them (and \textit{vice versa}).

The triangle inequality requires that, for any three feature allocations $ Z_1,Z_2, Z_3 $ in the space of all possible feature allocations, $ L(Z_1, Z_2) + L(Z_2, Z_3) \geq L(Z_1, Z_3)$. The triangle inequality establishes that if two feature allocations are similar to a third one, then those two feature allocations should also be similar to each other \citep{dahl2021search}. It also allows us to place an upper bound on the distance between two feature allocations if we know the distance from both of those feature allocations to another, meaning there is an intuitive structure to the space. In proving the triangle inequality for FARO loss, the following theorem (proven in the appendix) is useful.

\newtheorem*{thm:max_of_multiple_distances}{Theorem \ref{thm:max_of_multiple_distances}}

\begin{thm:max_of_multiple_distances}
The maximum of a finite set of functions that obey the triangle inequality also obeys the triangle inequality.
\end{thm:max_of_multiple_distances}

We will use Theorem \ref{thm:max_of_multiple_distances} to prove the triangle inequality for FARO loss. Recall the characterization of FARO loss $ L(Z, \hat{Z}) = \text{min} \{ L_1(Z, \hat{Z}), L_2(Z, \hat{Z}), \ldots, L_m(Z, \hat{Z}) \} $. Now, consider the function: \vspace{-0.35cm} $$ L^\star(Z, \hat{Z}) = \text{max} \{ L_1(Z, 1-\hat{Z}), L_2(Z, 1-\hat{Z}), \ldots, L_m(Z, 1-\hat{Z}) \}, \vspace{-0,5cm} $$ where $L^\star$ is a related function to FARO loss using the same generalized Hamming distance metrics $L_1, L_2, \ldots, L_m$, except now the \textit{maximum} of these values is selected, and the second argument in each loss function is $1-\hat{Z}$ (i.e.,  a matrix where each element of $\hat{Z}$ is ``flipped'' from a 1 to a 0 and \textit{vice versa}). By Theorem \ref{thm:max_of_multiple_distances} we have that $L^\star$ satisfies the triangle inequality. Further, the maximizer of $L^\star$ corresponds to the minimizer of the FARO loss $L$ (proven in the appendix). Therefore, the triangle inequality holds for FARO loss.

We conclude that FARO loss induces a quasi-metric space on feature allocations which behaves according to intuition for a distance function, with the exception of symmetry. However, our software maintains a standardized direction of computation such that asymmetry should not be a concern for users; they need only specify the value of $a$. When $a=b$, the symmetry property also holds, so in this special case, FARO loss is a metric and not just a quasi-metric (proven in appendix).

\section{FANGS Algorithm}
\label{sec:fangs}
Having established a loss function that can easily compute the distance between
feature allocations, we proceed to explain our novel search algorithm and how it
builds on the FARO loss function. We call our search algorithm FANGS, an acronym
for \textbf{f}eature \textbf{a}llocation \textbf{n}eighborhood-based
\textbf{g}reedy \textbf{s}earch.

\subsection{Algorithmic Description}

The FANGS algorithm takes a list of posterior feature allocation samples as its
main input, plus several tuning parameters. We introduce our description of this
search algorithm with pseudocode (in Algorithm \ref{algorithm:fangs}) and then
outline the purpose of each parameter. The search algorithm is implemented in
the \texttt{fangs} package in \texttt{R}.

\begin{algorithm}[tb]
	\small
	\setstretch{1}
	\caption{\small Pseudocode for the FANGS algorithm\label{fangsAlgorithm}.  Let $Z_1,\ldots,Z_B$ be samples from the feature allocation distribution of interest.
		Let $N_\text{iter}$ be the number of iterations, $K_\text{max}$ be the maximum number of features present among the samples, $N_\text{init}$ be the number of baseline samples used to obtain initial estimates, and $N_\text{sweet}$ be the number of initial estimates used as starting points for the ``sweetening phase'' (i.e., optimization). }
	\label{algorithm:fangs}
	\begin{algorithmic}[1]
		\State \Comment{Initialization phase}
		\State Augment samples with columns of zeros such that each has exactly $K_\text{max}$ features.
		\For{$i$ in 1,\ldots,$N_\text{init}$}
   		\State Let $Z^{(i)}$ be randomly sampled without replacement from $Z_1, \ldots, Z_B$.
   		\State Align the columns of the remaining $B-1$ samples to $Z^{(i)}$ to minimize Hamming distance.
   		\State Compute the element-wise means (proportions of ones) across the $B$ aligned samples.
   		\State Form $Z^*_i$ by thresholding all proportions (to 0 or 1) given the cutoff $\nicefrac{a}{2}$.
   		\State Remove any all-zero columns from $Z^*_i$.
		\State Calculate the expected FARO loss of $Z^*_i$.
		\EndFor
        \State Let $\zeta_1, \ldots, \zeta_{N_\text{sweet}}$ be the $Z^*$'s with the smallest expected FARO loss.
   		\State \Comment{Sweetening phase}
   		\For{$i$ in $1,...,N_{\text{iter}} $}
		\For{$j$ in $1,...,N_{\text{sweet}} $}
		\State Propose $\zeta_j^*$ by flipping a randomly selected entry of $\zeta_j$ from 0 to 1 or \textit{vice versa}.
		\If{the expected FARO loss of $\zeta_j^*$ is less than that of $ \zeta_j $}
		\State Set $ \zeta_j $ to $\zeta_j^*$.
		\Else
		\State Leave $\zeta_j$ unchanged.
		\EndIf
		\EndFor
		\EndFor
		\State \Return the feature allocation with the smallest expected FARO loss among $\zeta_1, \ldots, \zeta_{N_\text{sweet}}$.
	\end{algorithmic}
\end{algorithm}

FANGS is comprised of two stages: an initialization phase and a sweetening
phase. In the initialization phase, $N_{\text{init}}$ feature allocations are
randomly selected from the $B$ samples.  We refer to these $N_{\text{init}}$
feature allocations as baselines. For each baseline, the $B$ samples are
realigned by column to minimize their generalized Hamming distance to the
baseline. This is similar to the approach in \citet{zeng2018}, but our approach
allows the incorporation of differential Hamming distance penalties and is much
faster due to recognition of the linear assignment problem. This baseline
alignment approach is also different from their alignment approach which aligned
samples to each other sequentially, one after another, and is thus not
parallelizable. For each set of $B$ realigned samples, the element-wise mean is
calculated, yielding an $n \times K_\text{max}$ matrix of proportions. Each
proportions matrix is then thresholded to $0$ or $1$ using a cutoff point of
$\nicefrac{a}{2} \in (0,1)$. Lower values of $a$ result in more dense initial
estimates (i.e., more values are thresholded to 1) and higher $a$ leads to the
opposite. This concept is further discussed and verified in Section
\ref{subsec:diffA}. The result of the initialization phase is $N_{\text{init}}$
initial feature allocation estimates.

From these initial estimates, only the $N_{\text{sweet}}$ feature allocations
with the lowest expected FARO losses are advanced to the sweetening phase.
During the sweetening phase, the algorithm iterates through each of the
$N_{\text{sweet}}$ samples $N_{\text{iter}}$ times. At each iteration, a
randomly selected element of the binary matrix is ``flipped'' (i.e., switched
from 0 to 1 or 1 to 0) and the algorithm checks to see if the flip lowered the
expected loss. If so, the change is accepted and the old state is discarded;
otherwise, the matrix remains unchanged. After sweetening each of the
$N_{\text{sweet}}$ matrices $N_{\text{iter}}$ times, the feature allocation
having the lowest (Monte Carlo estimate of the) expected FARO loss is returned
as the point estimate.

\subsection{Discussion of Parameters}

\begin{figure}[tb]
    \centering
    \includegraphics[width=0.45\textwidth]{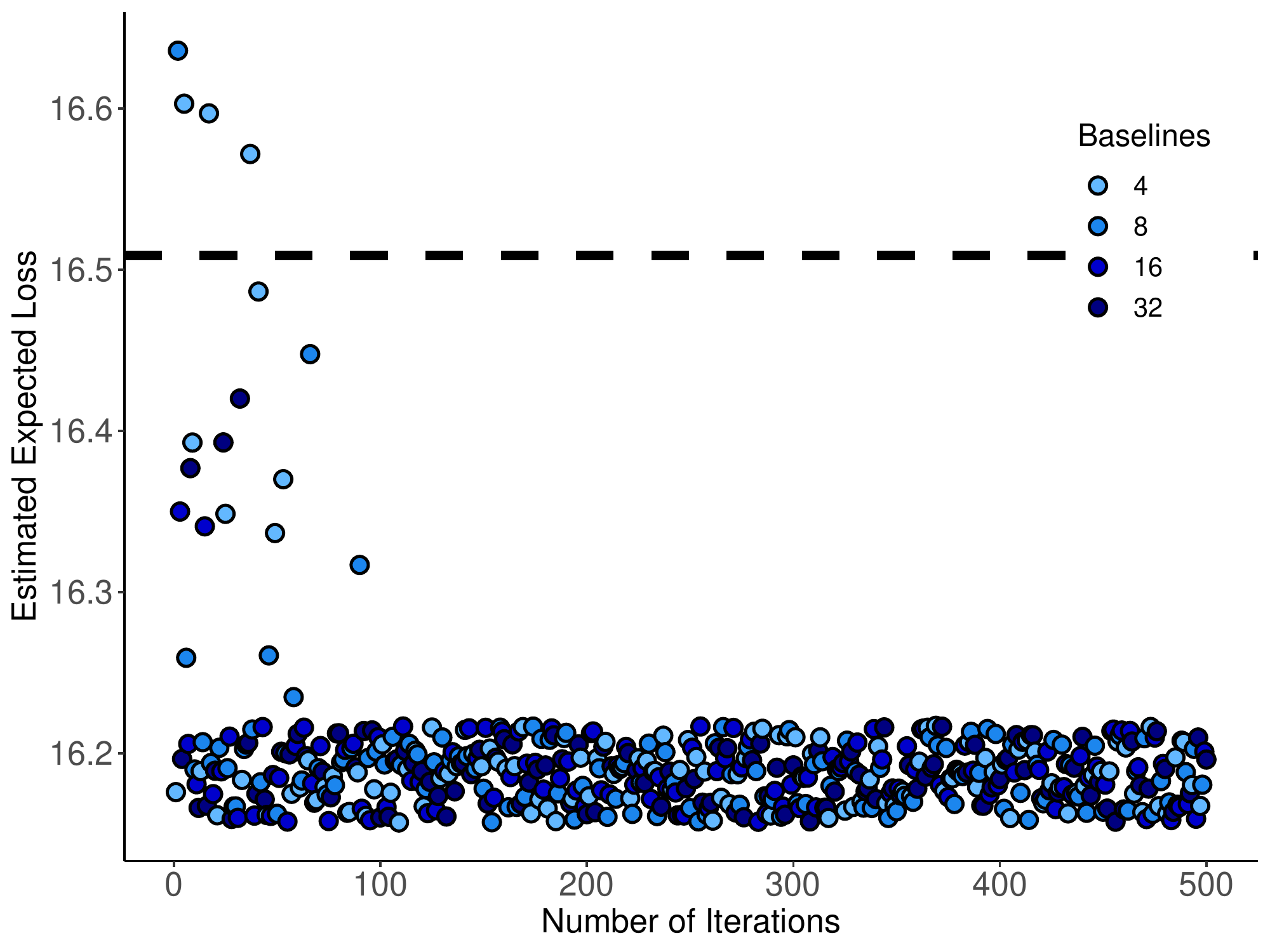}
    \hspace{1cm}
    \includegraphics[width=0.45\textwidth]{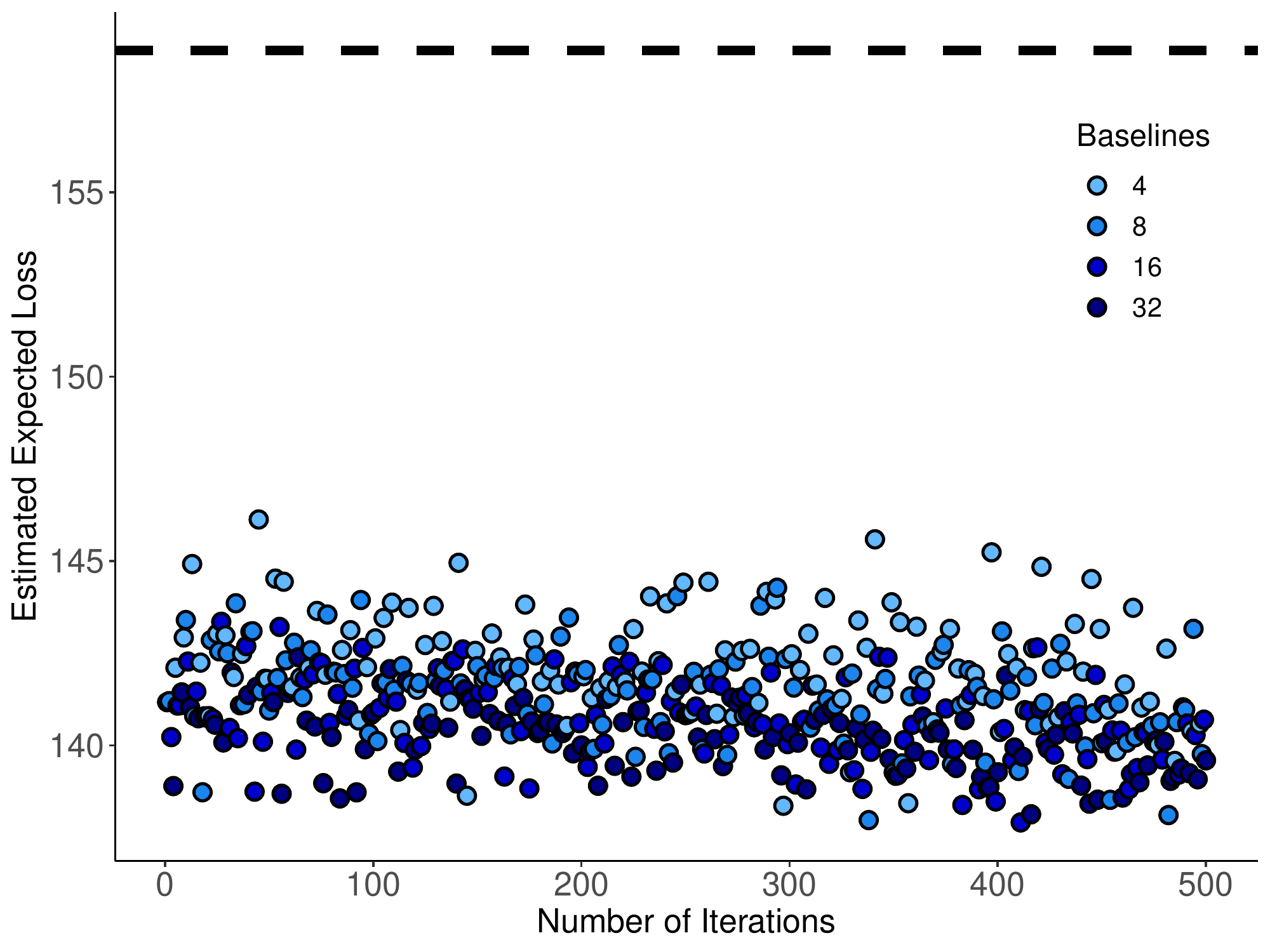}
    \caption{Jittered scatter plots displaying the estimated expected loss as a
    function of $N_{\text{iter}}$ for two sets of samples from
    \citet{warr2021attraction}. The dashed line indicates the estimated expected
    loss from the minimizing draw.}
    \label{fig:nIter}
\end{figure}

For most of the search's tuning parameters, there are versatile default values
that are sufficient for the vast majority of applications. The $N_{\text{iter}}$
argument is an influential parameter in terms of both loss minimization and wall
time. Raising the number of iterations tends to lead to estimates with lower
expected loss but also increases the computation time. The first set of samples
in Figure \ref{fig:nIter} was smaller in dimension ($n=20$, $K=3$), so it did
not take many iterations to stabilize to an estimate. The second set was larger
($n=62, K=77$) and continued to slowly find a better estimate given more
iterations. We suggest $N_{\text{iter}}=1000$ as the default value, which seems
to provide a balance between computation time and loss minimization for a
moderate number of samples and moderate matrix dimensions. However, for any
particular dataset and application, we suggest considering other values for
$N_{\text{iter}}$ to balance stability of the estimate and time.

The default values for $N_{\text{init}}$ and $N_{\text{sweet}}$ are 16 and 4,
respectively. These choices arise from the fact that both phases of FANGS (the
initialization and the sweetening) can be run in parallel. It is now common for
most systems, even personal desktops and laptops, to have 4 or more cores
available. Thus, it seems logical to allow a default of 4 initial estimates and
4 sweetenings. However, the baseline alignment is much faster than the
sweetening, and we found it helpful to use more baselines when possible. With
these parameters in mind, we offer Figure \ref{fig:fangs-def} as an illustration
of FANGS under its default values.

\begin{figure}[tb]
    \begin{tikzpicture}[
            draws/.style={circle, draw=blue!40, fill=blue!3, very thick, minimum size=12mm},
            candidates/.style={circle, draw=gray!90, fill=gray!25, very thick, minimum size=12mm},
            intersect/.style={circle, draw=black, fill=black, inner sep=0pt, minimum size=4mm},
            steps/.style={text width=6.9cm},
            box/.style={rectangle, draw, minimum height=4cm, minimum width=8.25cm},
            ]
        \node[draws] (z1) at (0,0) {$Z_1$};
        \node[draws] (z2) [right=.4cm of z1] {$Z_2$};
        \node[draws] (z3) [right=.4cm of z2] {$Z_3$};
        \node[] (zdot) [right=.4cm of z3] {\ldots};
        \node[draws] (zn) [right=.4cm of zdot] {$Z_B$};
        \node[steps] (step1) [left=.4cm of z1] {\footnotesize A) Randomly select 16 baselines from the $B$ samples: };
        
        \node[intersect] (temp) at (2.9,-1.25) { };
        \node[] (intertop) at (2.9,-1.2) {};
        \node[] (interbottom) at (2.9,-1.3) {};
        
        \node[candidates] (c1) [below=1.5cm of z1] {$Z^{(1)}$};
        \node[candidates] (c2) [below=1.5cm of z2] {$Z^{(2)}$};
        \node[candidates] (c3) [below=1.5cm of z3] {$Z^{(3)}$};
        \node[] (cdot) [right=.4cm of c3] {\ldots};
        \node[candidates] (cn) [below=1.5cm of zn] {$Z^{(16)}$};
        \node[steps] (step2) [left=.4cm of c1] {\footnotesize B) Align all $B$ samples to each baseline, compute proportions, and threshold: };
        
        \node[draws] (star1) [below=.6cm of c1] {$Z^*_1$};
        \node[draws] (star2) [below=.6cm of c2] {$Z^*_2$};
        \node[draws] (star3) [below=.6cm of c3] {$Z^*_3$};
        \node[] (stardot) [right=.4cm of star3] {\ldots};
        \node[draws] (starn) [below=.6cm of cn] {$Z^*_{16}$};
        
        \node[candidates] (s1) [below=2.2cm of star1] {\large $\zeta_j$};
        \node[steps] (step3) [below=1.2cm of step2] {\footnotesize C) Score the expected loss for each initial estimate and keep the 4 best for sweetening: };
        
        \node[candidates] (s2) [below=2.2cm of starn] {\large $\zeta^\star_j$};
        \node[steps] (step4) [left=0.4 cm of s1] {\footnotesize D) Randomly flip entries 1,000 times for each sweetening: };
        
        \node[steps] (step5) [below=0.6cm of step4] {\footnotesize E) Final estimate $\boldsymbol{\zeta}$ is the $\zeta^\star_j$ with the lowest expected loss.};
        \node[box] (r) at (3.1,-8.2) { };
        \node[] (enterbox) [below=0.75cm of star1] {};
        \node[] (enterbox2) [right=0.1mm of enterbox] {};
        \node[] (enterbox3) [right=0.1mm of enterbox2] {};
        \node[] (exitbox) [below=1.15cm of s2] {};
        \node[steps] (nbests) at (2.6, -10) {\footnotesize For $ j = 1, 2, 3, 4 $, and 1,000 times for each $j$};

        \path[-] (z1) edge [bend right=15] (intertop);
        \path[-] (z2) edge [bend right=15] (intertop);
        \path[-] (z3) edge [bend left=15] (intertop);
        \path[-] (zn) edge [bend left=15] (intertop);
        \path (interbottom) edge [bend right=15] (c1);
        \path (interbottom) edge [bend right=15] (c2);
        \path (interbottom) edge [bend left=15] (c3);
        \path (interbottom) edge [bend left=15] (cn);
        \path (c1) edge (star1);
        \path (c2) edge (star2);
        \path (c3) edge (star3);
        \path (cn) edge (starn);
        \path (star1) edge (enterbox);
        \path (star2) edge [bend left=15] (enterbox2);
        \path (star3) edge [bend left=15] (enterbox3);
        \path (s1) edge [bend left=30] node[above,pos=.5] {\footnotesize Keep change if expected loss decreases} (s2);
        \path (s2) edge [bend left=30] node[below,pos=.5] {\footnotesize Revert otherwise} (s1);
        
        \node[] (turn1) [below=0.4cm of exitbox] {};
        \node[] (turn2) [left=11.8cm of turn1] {};
        \node[] (turn3) [above=0.3cm of turn2] {};
        \draw (exitbox) -- (turn1.center) -- (turn2.center) -- (turn3);

    \end{tikzpicture}
    \vspace{4pt}
    \caption{FANGS algorithm overview using default settings.}
    \label{fig:fangs-def}
\end{figure}

\subsection{Computational Complexity}

We already established the computational complexity for our FARO loss
computation, derived from the J\"onker-Volgenant Algorithm and implemented in
Rust \citep{pkg.lapjv}, to be $\mathcal{O}(K^3)$ for comparing feature
allocations having up to $K$ columns. To examine the computational burden of
FANGS as a whole, we frame this $\mathcal{O}(K^3)$ complexity in the context of
repeatedly scoring the expected loss for many feature allocations among a set of
$B$ posterior samples.

To obtain $N_\text{init}$ sets of realigned samples, FARO loss is computed $B \,
N_\text{init}$ times, yielding a complexity of $\mathcal{O}(K^3 \, B \,
N_\text{init})$. Then, the expected loss for each of the $N_\text{init}$ initial
estimates is computed, which also has complexity $\mathcal{O}(K^3 \, B \,
N_\text{init})$. Adding these components and simplifying yields $\mathcal{O}(K^3
\, B \, N_\text{init} + K^3 \, B \, N_\text{init}) \equiv \mathcal{O}(2K^3 \, B
\, N_\text{init}) \equiv \mathcal{O}(K^3 \, B \, N_\text{init})$ in the
initialization phase. In the sweetening phase, the expected FARO loss is
estimated once per iteration in each sweetening. Thus, the order of the
sweetening phase is $\mathcal{O}(K^3 \, B \, N_{\text{iter}} \,
N_{\text{sweet}})$. These two phases are additive in FANGS, so the order of the
search as a whole is $\mathcal{O}(K^3 \, B \, N_\text{init} + K^3 \, B \,
N_{\text{iter}} \, N_{\text{sweet}})$, or equivalently, $\mathcal{O} \Big( (K^3
\, B ) \, (N_\text{init} + N_{\text{iter}} \, N_{\text{sweet}}) \Big)$. Although
this may seem convoluted, the key takeaway is that the number of features $K$ is
the most influential in adding to or lightening the computational expense of
FANGS since it has the highest degree at 3. $K$ may be slightly different for
each matrix comparison since the number of features varies across draws, but if
this variance is low, then the order could still be approximated as above using
$K=\hat{K}$, the modal number of features from the samples.

\section{Verifications}
\label{sec:verifications}
In this section, we compare the FANGS estimate to the minimizing draw for five
different sets of posterior feature allocation samples obtained from previous
studies.

\subsection{Description of Posterior Samples}

Of the five sets of posterior feature allocation samples examined, two of them
are from \citet{warr2021attraction} and were already used in Section
\ref{sec:fangs} (a smaller set simulated directly from their linear latent
feature model and a larger set from real Alzheimer's disease data). In addition,
we have samples from the CMC study \citep{ni-cmc}, the SIFA study
\citep{zeng2018}, and scenario 1 of the DFA model
\citep{ni2020bayesian}\footnote{In the DFA study, $Z$ followed a complex
structure with 6 features. Some information in the first two features was
treated as fixed, so we ignore these first two features and assume $K=4$.}.
These papers were discussed previously and their posterior samples are
summarized in Table \ref{tab:dimensions}.

\begin{table}[tb]
    \centering
    \begin{tabular}{p{56mm}rrr}
        \toprule[1.5pt]
        Case Study & Observations ($n$) & Features ($K$) & Samples ($B$)  \\
        \midrule
        DFA \citep{ni2020bayesian} & 300 & 4 & 500 \\
        CMC \citep{ni-cmc} & 800 & 5 & 500 \\
        SIFA \citep{zeng2018} & 80 & 6 & 4000 \\
        AIBD \citep{warr2021attraction} & 20 & 3 & 1000 \\
        Alzheimer's \citep{warr2021attraction} & 62 & 77 & 3000 \\
        \bottomrule[1.5pt]
    \end{tabular}
    \bigskip
    \caption{Values of $n$ and $K$ for the true feature allocation matrix $Z$
    that was used to generate $B$ samples in the simulations for each case
    study. The Alzheimer's samples were based on real data, so 77 represents
    $\hat{K}$ instead of the ``true $K$''. }
    \label{tab:dimensions}
\end{table}

\subsection{Simulation Study}

It only makes sense to compare search algorithms if the loss function is held
constant across all searches. Having already outlined why FARO loss holds
practical advantages over the others, we fix FARO loss as the loss function for
all case studies and compare the FANGS estimate against the loss-minimizing
draw. The simulations were run on a server with 512 GB of RAM and two Intel Xeon
Gold 6142 CPU @ 2.60GHz processor, yielding 32 physical cores and 64 cpu
threads. Although FANGS is a stochastic algorithm, the use of 16 baselines and
1,000 iterations resulted in almost no noticeable stochasticity. Table
\ref{tab:exp-losses} shows the estimated expected loss and wall time under each
estimate and case study. Not only does FANGS find estimates with lower expected
loss than the minimizing draw, but it appears to do so in a shorter window of
time when $K$ is higher. FANGS takes longer for the AIBD and DFA studies, is
about the same for the CMC study, and is much faster for the SIFA and
Alzheimer's samples.

\begin{table}[tb]
\centering
    \begin{tabular}{l@{\hskip 0.4cm}rrrrr}
        \toprule[1.5pt]
        & \multicolumn{2}{c}{\textbf{Draws Method}} & & \multicolumn{2}{c}{\textbf{FANGS}}  \\
        \cmidrule{2-3} \cmidrule{5-6}
        Study & Exp. Loss & Runtime & & Exp. Loss & Runtime \\
        \midrule
        DFA  & 17.51  & 0.14 & & 14.78  & 0.36 \\ 
        CMC  & 103.35 & 0.70 & & 102.21 & 0.72 \\
        SIFA & 50.45  & 3.77 & & 43.48  & 1.83 \\
        AIBD & 16.51  & 0.21 & & 16.19  & 0.70 \\ 
        Alzheimer's & 158.84 & 151.18 & & 138.94 & 32.06 \\
        \bottomrule[1.5pt]
    \end{tabular}
    \bigskip
    \caption{The expected loss from the minimizing draw and from the FANGS
    estimate (using default parameters), as well as the corresponding wall times
    (in seconds) averaged over 1000 replications for each case study. Monte
    Carlo error was nonexistent for the first four case studies and negligible
    for the Alzheimer's samples.}    
    \label{tab:exp-losses}
\end{table}

In short, FANGS reliably produces a feature allocation estimate with
significantly lower expected loss than the draws estimate in a comparable window
time (if not more quickly) using default parameters. FARO loss and FANGS are
implemented in \texttt{R} with a fast and easy-to-use software package. FANGS
could also be applied to other loss functions should others be developed in the
future. It is critical, however, that any new loss function be able to be
rapidly and repeatedly evaluated, otherwise it may not be viable in a search
algorithm like FANGS.

\subsection{Differential Penalties}
\label{subsec:diffA}

In Section \ref{subsec:diff_penalties_hamming}, we introduced generalized
Hamming distance and discussed using $a \neq b$ to assign lower FARO loss to
more dense or sparse feature allocations. As a test case, we examine the
Alzheimer's samples from the AIBD study ($n=62$), which had samples ranging from
$K=5$ to $K=94$. Figure \ref{fig:diffA} shows that the feature allocation
estimate is more dense when $a$ is lower (closer to 0) and more sparse as $a$
approaches 2. Even when $K$ didn't change by as much between increments of $a$
(i.e. $a=0.1$ and $a=0.5$), the sparsity of the individual features still did
(from 790 to 446 total entries in that instance). Furthermore, the largest
decreases in $K$ (for the estimates) correspond to the highest-density region of
the distribution of $K$ in the original samples. Similar trends were present for
the other sets of samples, although it is more prominent here due to the high
variance of $K$. We conclude that the incorporation of differential penalties in
FARO loss (and by extension, FANGS) successfully allows for sparsity tuning in
the feature allocation estimation process.

\begin{table}
	\begin{minipage}{5cm}
	\centering
	\begingroup
    \renewcommand{\arraystretch}{1.01} 
	\begin{tabular}{llrr}
        \toprule[1.5pt]
          $a$ & $b$ & $K$ & Entries \\
          \midrule
          0.01 & 1.99 & 85 & 1722 \\
          0.1  & 1.9  & 81 & 790  \\
          0.5  & 1.5  & 79 & 446  \\
          1.0  & 1.0  & 74 & 338  \\
          1.5  & 0.5  & 60 & 242  \\
          1.9  & 0.1  & 27 & 112  \\
          1.99 & 0.01 & 6  & 14   \\
        \bottomrule[1.5pt]
    \end{tabular}
    \endgroup
	\end{minipage}
	\begin{minipage}{11cm}
		\centering
		\includegraphics[scale=0.42]{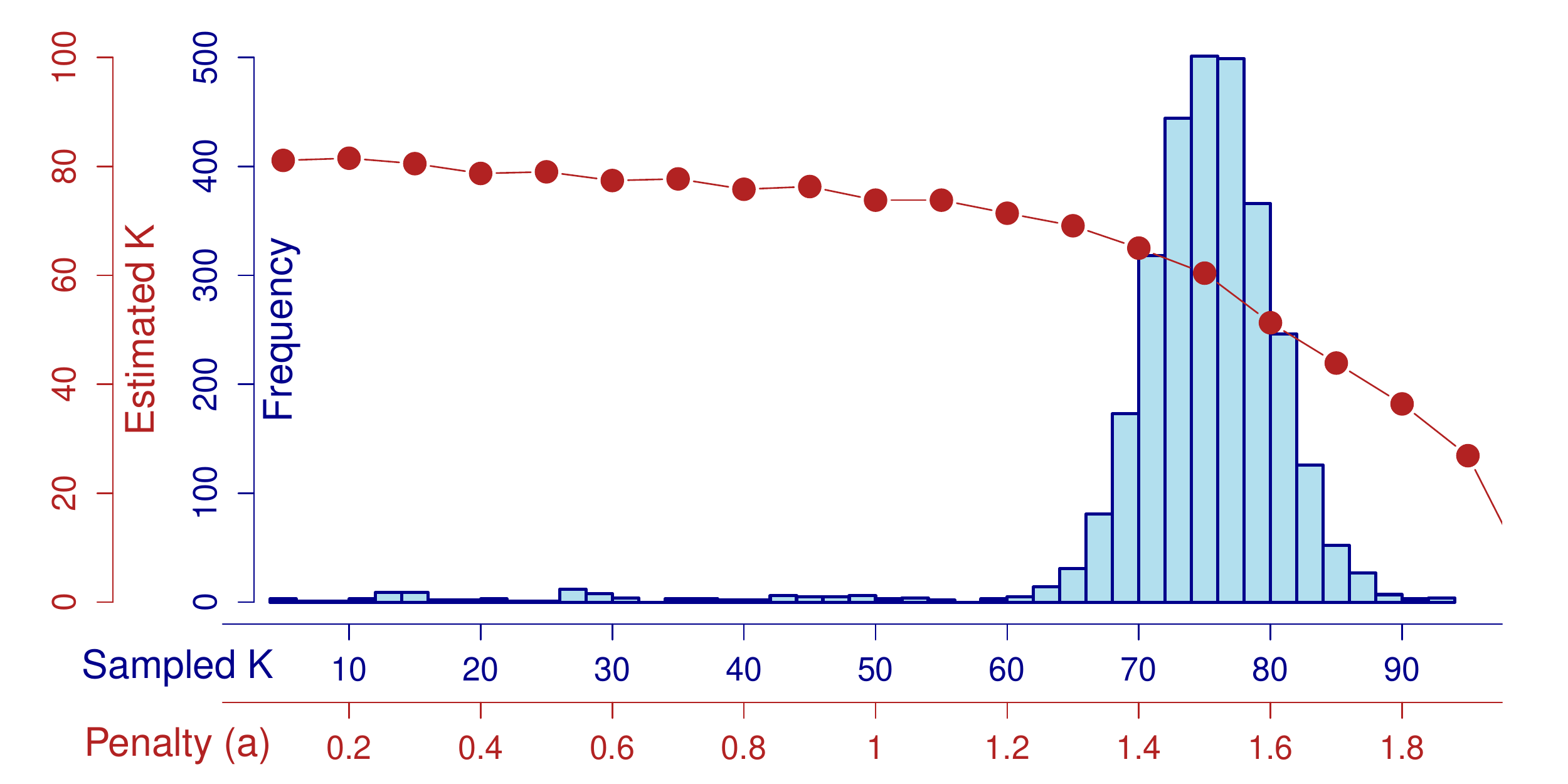}
	\end{minipage}
	\captionof{figure}{Total number of features $K$ and entries in the FANGS
	estimate from the Alzheimer's samples for $a \in (0,2)$, overlaid with the
	distribution of $K$ in the original samples.}
	\label{fig:diffA}
\end{table}

\section{Conclusion}
\label{sec:conclusion}
Previous feature allocation studies typically developed their own loss function
tailored directly to the estimation problem at hand, then performed minimization
over the posterior samples. FARO loss is more broadly applicable, accommodates
feature allocations with different numbers of features, and does not depend on
the feature ordering. It satisfies the properties of a quasi-metric and is
computed quickly using optimization techniques from the linear assignment
problem. Because the loss is so rapidly evaluated, FANGS is able to better
explore the feature allocation space than the draws method. In the case studies
we examined, FANGS consistently found an estimate with lower expected loss than
any of the posterior samples. We believe that FARO loss and FANGS are effective
and versatile enough to serve as a general loss function and search algorithm
for feature allocation estimation problems.

\section{Appendix}
\label{sec:appendix}
\begin{thm}\label{thm:max_of_multiple_distances}
The maximum of a finite set of functions that obey the triangle inequality also
obeys the triangle inequality.
\end{thm}

\noindent \textbf{Proof:} Let $X$ be a set of all feature allocations for $n$
items and let $d_1, \ldots, d_m$ be real-valued functions defined on $X \times
X$ which obey the triangle inequality. Without loss of generality, we first seek
to show that the maximum of $d_1$ and $d_2$ obeys the triangle inequality. For
all $x,y\in X$, let $d_{1,2}(x,y) = \max\{d_1(x,y), d_2(x,y)\}$. We must show
that, for any $x,y,z\in X$, $ d_{1,2}(x,y) + d_{1,2}(y,z) \geq d_{1,2}(x,z)$.
Recall that, since both $d_1$ and $d_2$ obey the triangle inequality, $d_1(x,y)
+ d_1(y,z) \geq d_1(x,z)$ and $d_2(x,y) + d_2(y,z) \geq d_2(x,z)$.  But, by the
definition of $d_{1,2}$, the left-hand-side of each of these inequalities is
less than or equal to $ d_{1,2}(x,y) + d_{1,2}(y,z) $. Then, $ d_{1,2}(x,y) +
d_{1,2}(y,z) $ is greater or equal to both $d_1(x,z)$ and $d_2(x,z)$ and,
therefore, greater than or equal to $\max\{ d_1(x,z), d_2(x,z)\} =
d_{1,2}(x,z)$. Therefore $d_{1,2}$ obeys the triangle inequality.

We can again apply this result to get that $d_{1,2,3}(x,y) = \max
\{d_{1,2}(x,y), d_3(x,y)\} = \max \{ \max \{ d_1(x,y), d_2(x,y) \}, d_3(x,y)\} =
\max \{d_1(x,y), d_2(x,y), d_3(x,y)\}$ also obeys the triangle inequality.
Repeatedly applying this result for all integers until $m$ yields:
\vspace{-0.35cm}
$$  d_{1,\cdots,m}(x,y) = \max \{d_{1,\cdots,m-1}(x,y), d_m(x,y)\} = \max \{d_1(x,y), \cdots, d_m(x,y)\} \vspace{-0.35cm} $$ 
Hence, $d_{1,\cdots,m}(x,y)$ obeys the triangle inequality.

\begin{thm}\label{thm:gen_hamming_quasi}
    The generalized Hamming distance is a quasi-metric on the feature allocation
    space for any $a,b>0$ where $a+b=2$.
\end{thm}

\noindent For brevity, we denote $\mathds{I}(Z_{ij}=1)$ as $\mathbb{Z}_{ij}$ and
$\mathds{I}(Z_{ij}=0)$ as $\mathbb{Z}_{-ij}$ for any binary feature allocation
matrix $Z$. Also, let $\mathds{I}(Z_{ij}=1 \cap \hat{Z}_{ij}=1) =
\mathbb{Z}_{ij} \, \mathbb{\hat{Z}}_{ij} $.

\noindent \textbf{Proof:}
\textit{Identity of Indiscernibles:} Suppose $Z=\hat{Z}$. Then for all $1<i<n, 1<j<K$, we have that $ \mathbb{Z}_{ij} \, \mathbb{\hat{Z}}_{-ij} = 0 $ and $ \mathbb{Z}_{-ij} \, \mathbb{\hat{Z}}_{ij} = 0 $. Then, $\sum_{i=1}^n \sum_{j=1}^K [ a(0) + b(0) ] = 0 $, so $L(Z,\hat{Z})=0$. We prove the converse using its contrapositive. Suppose $Z\neq \hat{Z}$; that is, there exists at least one pair of $(i,j)$ such that $Z_{ij} \neq \hat{Z}_{ij}$. It follows that: \vspace{-0.35cm} $$ L(Z,\hat{Z}) = \sum_{i=1}^n \sum_{j=1}^K \Big[ a \, \mathbb{Z}_{ij} \, \mathbb{\hat{Z}}_{-ij} + b \, \mathbb{Z}_{-ij} \, \mathbb{\hat{Z}}_{ij} \Big] \geq \min{(a,b)} \vspace{-0.35cm} $$
Since $a$ and $b$ are both positive real numbers, $\min{(a,b)}>0$, and so $L(Z,\hat{Z})>0$.

\textit{Triangle Inequality:} Given three feature allocations $U, V, W$ defined
on the same feature allocation space, the triangle inequality holds for
generalized Hamming distance if $L(U,V) + L(V,W) - L(U,W) \geq 0$. The left side
of this expression simplifies to:
$$ a \cdot \sum_{i=1}^n \sum_{j=1}^K \Big[ \mathbb{U}_{ij} \, \mathbb{V}_{-ij} + \mathbb{V}_{ij} \, \mathbb{W}_{-ij} - \mathbb{U}_{ij} \, \mathbb{W}_{-ij} \Big] +  b \cdot \sum_{i=1}^n \sum_{j=1}^K \Big[ \mathbb{U}_{-ij} \, \mathbb{V}_{ij} + \mathbb{V}_{-ij} \, \mathbb{W}_{ij} - \mathbb{U}_{-ij} \, \mathbb{W}_{ij}  \Big] $$
We examine each of the two summations to see if either component can be less
than zero for any $(i,j)$ index. We can also disregard $a$ and $b$ for this
purpose since they are both positive real values. The only way the first
summation can be less than zero is if $ \mathbb{U}_{ij} \, \mathbb{V}_{-ij} $
and $ \mathbb{V}_{ij} \, \mathbb{W}_{-ij} $ are both false, while $
\mathbb{U}_{ij} \, \mathbb{W}_{-ij} $ also holds true. For the first two
indicators to both be false, we need one of the following conditions to hold:
\begin{multicols}{4}
\begin{enumerate}[(i)]
    \item $\mathbb{U}_{-ij} \, \mathbb{V}_{-ij} \, \mathbb{W}_{-ij}$ 
\end{enumerate}
\columnbreak
\begin{enumerate}[(i)]
    \setcounter{enumi}{1}
    \item $\mathbb{U}_{ij} \, \mathbb{V}_{ij} \, \mathbb{W}_{ij}$ 
\end{enumerate}
\columnbreak
\begin{enumerate}[(i)]
    \setcounter{enumi}{2}
    \item $\mathbb{U}_{-ij} \, \mathbb{V}_{ij} \, \mathbb{W}_{ij}$
\end{enumerate}
\columnbreak
\begin{enumerate}[(i)]
    \setcounter{enumi}{3}
    \item $\mathbb{U}_{-ij} \, \mathbb{V}_{-ij} \, \mathbb{W}_{ij}$
\end{enumerate} 
\end{multicols}
\noindent In any of the above cases, the indicator $ \mathbb{U}_{ij} \,
\mathbb{W}_{-ij} $ does not hold true, and so the first line of the expression
would evaluate to zero. A nearly identical approach follows for the second
summation with $b$. Finally, we conclude that $L(U,V) + L(V,W) - L(U,W) \geq 0$
because it is a sum of strictly nonnegative terms. Therefore $L(U,V) + L(V,W)
\geq L(U,W)$, and the triangle inequality is met for generalized Hamming
distance.

\begin{lemma}\label{lemma:max_min_FARO}
	The maximizing column permutation of $L^\star$ corresponds to the minimizing
	permutation of FARO loss.
\end{lemma}

\noindent \textbf{Proof:}
The following proof holds regardless of whether $a$ and $b$ are equal or not. Recall the function $L^\star$ which was previously defined in Section \ref{sec:new_loss} as: \vspace{-0.35cm} $$ L^\star(Z,\hat{Z}) = \text{max}\{ L_1(Z,1-\hat{Z}), L_2(Z,1-\hat{Z}), \ldots, L_m(Z,1-\hat{Z}) \}, \text{ where } m=K! \vspace{-0.35cm} $$ Denote $L_i(Z, 1-\hat{Z})$ as $L(Z, 1-\hat{Z}^{(i)})$, where $L$ is just the generalized Hamming distance function but now $\hat{Z}^{(i)}$ is the feature allocation matrix $\hat{Z}$ with columns permuted according to the $i$th permutation, $i = 1,2,\ldots,m$. Then $L(Z, 1-\hat{Z})$ is simplified as follows:

\begin{align*}
    &\sum_{\ell=1}^n \sum_{j=1}^K \Big[ a \, \mathbb{Z}_{\ell j} \, (1-\hat{Z}^{(i)})_{- \ell j} + b \, \mathbb{Z}_{-\ell j} \, (1-\hat{Z}^{(i)})_{\ell j} \Big] = \sum_{\ell=1}^n \sum_{j=1}^K \Big[ a \, \mathbb{Z}_{\ell j} \, \hat{Z}^{(i)}_{\ell j} + b \, \mathbb{Z}_{-\ell j} \, \hat{Z}^{(i)}_{-\ell j} \Big] \\
    &= nK(a+b) - \sum_{\ell=1}^n \sum_{j=1}^K \big[ a \, \mathbb{Z}_{-\ell j} + b \, \mathbb{Z}_{\ell j} \big] - \big[ a \cdot \sum_{\ell=1}^n \sum_{j=1}^K \mathbb{Z}_{\ell j} \, \hat{Z}^{(i)}_{-\ell j} + b \cdot \sum_{\ell=1}^n \sum_{j=1}^K \mathbb{Z}_{-\ell j} \, \hat{Z}^{(i)}_{\ell j} \big] \\
    &= nK(a+b) - \sum_{\ell=1}^n \sum_{j=1}^K \big[ a \, \mathbb{Z}_{-\ell j} + b \, \mathbb{Z}_{\ell j} \big] - L(Z, \hat{Z}^{(i)})
\end{align*}
Note that the first two terms in the final line of the equality do not depend on the column ordering of $Z$ or $\hat{Z}$, since the total numbers of ones or zeros in $Z$ is the same no matter how its columns are arranged. Therefore, we have: \vspace{-0.35cm} $$ \text{arg\,max}_i \{ L(Z,1-\hat{Z}^{(i)}) \} = \text{arg\,max}_i \{- L(Z,\hat{Z}^{(i)}) \} = \text{arg\,min}_i \{ L(Z,\hat{Z}^{(i)}) \} \vspace{-0.35cm} $$

We have now shown that the maximizing column permutation for $L^\star$ is the
same as the minimizing column permutation for $L$, or FARO loss.

\begin{lemma}\label{lemma:symmetry_faro}
	The symmetry property holds for FARO loss in the special case that $a=b=1$.
\end{lemma}

\noindent \textbf{Proof:} The notation from Section \ref{sec:new_loss} is used.
Note that when $a=b=1$, $L_1, L_2, \ldots, L_m$ are all metrics, as the
generalized Hamming distance in this special case reverts to the normal Hamming
distance (which \citep{robinson2003} proved is a metric). A loss function $L$
satisfies symmetry when $ L(Z, \hat{Z}) = L(\hat{Z}, Z) \hspace{3pt}\forall
\hspace{3pt}Z,\hat{Z} $. It is simple to see that $ L(Z, \hat{Z}) = \text{min}
\{ L_1(Z, \hat{Z}), L_2(Z, \hat{Z}), \ldots, L_m(Z, \hat{Z}) \} = \text{min} \{
L_1(\hat{Z}, Z), L_2(\hat{Z}, Z), \ldots, L_m(\hat{Z}, Z) \} $ since \\ $L_1,
L_2, \ldots, L_m$ are all metrics. By definition of $L$, we have $\text{min} \{
L_1(\hat{Z}, Z), L_2(\hat{Z}, Z), \ldots, \\ L_m(\hat{Z}, Z) \} = L(\hat{Z}, Z)
$. Thus $L(Z, \hat{Z}) = L(\hat{Z},Z)$, so the symmetry property holds.

\bibliographystyle{asa}
\bibliography{references}

\end{document}